\documentclass[
UTF8,
reprint,
amsmath,amssymb,
aps
]{revtex4-2}
\usepackage[normalem]{ulem}
\usepackage{graphicx}
\usepackage{dcolumn}
\usepackage{bm}
\usepackage{hyperref}

\usepackage{subfigure}
\UseRawInputEncoding
\begin{document}

	\preprint{APS/123-QED}
	
	\title{Relieving the $S_8$ Tension: Exploring the Surface-type DBI Model as a Dark Matter Paradigm} 
	
	\author{Xingpao Suo}
	\email{xpsuo@zju.edu.cn}
	\affiliation{Institute for Astronomy, School of Physics, Zhejiang University, Hangzhou 310027, China}
	
	\author{Xi Kang}
	\email{kangxi@zju.edu.cn}
	\affiliation{Institute for Astronomy, School of Physics, Zhejiang University, Hangzhou 310027, China}
    \affiliation{Purple Mountain Observatory, 10 Yuan Hua Road, Nanjing 210034, China}

	\author{Huanyuan Shan}
	\email{hyshan@shao.ac.cn}
	\affiliation{Shanghai Astronomical Observatory (SHAO), Nandan Road 80, Shanghai 200030, China}

	\date{\today}
	
	\begin{abstract}

       Recent observations from weak gravitational lensing (WL) surveys indicate a smoother Universe compared to the predictions of the Cosmic Microwave Background (CMB). This inconsistency is commonly referred to as the $\sigma_8$ tension or $S_8$ tension, where $\sigma_8$ represents the present root-mean-square matter fluctuation averaged over a sphere of radius $8 h^{-1} \mathrm{Mpc}$, and $S_8 \equiv \sigma_8\sqrt{\Omega_m/0.3}$.  In this article, we investigate a kind of general Dirac-Born-Infeld (DBI)  Lagrangian referred to as \textit{surface-type DBI} (sDBI)  model. We find that up to the linear order, the constraints on the sDBI model with high-redshift probe (CMB) and low-redshift probes (WL and Galaxy Clustering, GC) yield $S_8= 0.7448_{-0.21}^{+0.031}$ and $0.7426_{-0.085}^{+0.054}$, respectively. Remarkably, these values not only demonstrate self-consistency but also align with the values obtained from the majority of low-redshift probes.
       Furthermore, we present a discussion on exploring the non-linear effects of this model, which holds the potential to address additional challenges associated with Cold Dark Matter (CDM) on small scales.
	\end{abstract}
	
	\maketitle
	
	
	
	\textit{Introduction.} --The $\Lambda$CDM model stands as the most widely accepted cosmological model, serving as the standard framework for Big Bang cosmology. While the $\Lambda$CDM model provides a straightforward and successful description that aligns with a wide range of observations, the advancement of theoretical and observational studies has brought to light certain inconsistencies. These disparities, whether arising from conflicts between different observations or discrepancies between theory and observations, have begun to challenge the $\Lambda$CDM model, indicating the necessity for new extended models or alternative physics\citep{ABDALLA202249}. Among the various challenges faced by the $\Lambda$CDM model, the issue of $\sigma_8$ or $S_8$ tension stands out as one of the most prominent\citep{DIVALENTINO2021102604}. It shows that the low-redshift probes such as Weak gravitational Lensing (WL) \citep{Asgari2021,PhysRevD.105.023520}, Galaxy Clustering (GC) \citep{Salvati2018, Ivanov_2020} as well as their combined analyses \citep{Corasaniti_2021,Heymans2021}, indicate a smoother Universe than the prediction by Cosmic Microwave Background (CMB)\citep{Planck2018}. Quantitatively, the structure growth parameter $S_8$ derived from low-redshift probes consistently shows a $2-3\sigma$ lower value compared to the value obtained from the CMB. \citep{Planck2018,DES2018,Hikage2019,Wright2020,Asgari2021,Heymans2021, ABDALLA202249, poulin2022sigma, DESY3,HSCY3, Nunes:2021ipq}. 
    Recently, a joint cosmological analysis of cosmic shear + galaxy-galaxy lensing + GC yielded a constraint of $(\Omega_m, S_8)  = (0.305^{+0.010}_{-0.015},0.766^{+0.020}_{-0.014})$(see \cite{Heymans2021}, hereafter referred as \textit{K1K-3$\times$2pt}), where $S_8$ is low by $8.3 \pm 2.6\%$ compared to $(\Omega_m, S_8) = (0.3166\pm0.0084, 0.834\pm0.016)$ given by baseline of \textit{Planck2018}\citep{Planck2018}.
	
	Some new models has been proposed to solve or relieve $S_8$ tension, such as an additional scaling parameter on the CMB lensing amplitude\citep{sym10110585,PhysRevD.92.121302}, a dark energy and dark matter interaction model \citep{DiValentino:2019ffd,LUCCA2021100899,Valeentino2020}, and modified gravitation\citep{Planck2015beyond}, most of which give a consistent result with both CMB and low-redshift probes. 
	
	In this article, we propose a novel dark matter model that provides an alternative framework to resolve the $S_8$ tension. Our model, referred to as the sDBI model, introduces an area functional form as the dark matter Lagrangian, representing a specific instance within the broader class of general DBI models. Our investigation showcases the efficacy of this model in alleviating the $S_8$ tension by attenuating the formation of structures at low redshifts while maintaining the accurate evolution of perturbations at high redshifts.
	
	\textit{The surface-type DBI as a dark matter model.} --Here we consider the Lagrangian 
	\begin{eqnarray}
		\mathcal{L} 
		 \equiv
		\frac{R}{2\kappa} + 
		\Lambda_{I} 
		+ \Lambda_{II} \sqrt{1 + \partial_\mu\phi \partial^\mu\phi } + 
		\mathcal L_m \  \label{equ:lag}
	\end{eqnarray}
    and its corresponding action 
    $S = \int d^4x \sqrt{-g} \mathcal{L}$,
    where $g \equiv \det(g_{\mu\nu})$ represents the determinant of the space-time metric $g_{\mu\nu}$ with signature $[-1,1,1,1]$, 
    $R$ denotes the scalar curvature of Levi-Civita connection, $\kappa \equiv 8 \pi G$ with gravitational constant $G$, 
    $ \Lambda_I$ is the vacuum energy or equivalently cosmological constant,
    $\mathcal L_m $ is the Lagrangian of normal matter including radiation and baryon, 
    and $\Lambda_{II}\sqrt{1 + \partial_\mu\phi \partial^\mu\phi } $ with a constant $\Lambda_{II}$ and scalar field $\phi$ is the Lagrangian that we introduce to represent dark matter, which we refer to as the surface-type Dirac-Born-Infeld (sDBI) model.  Note that the term \textit{surface-type} comes from a mathematical standpoint. The term $\int d^4 x \sqrt{-g} \sqrt{1 + \partial_\mu\phi \partial^\mu\phi}$ can be viewed as formal area functional, which is usually used to describe the area of a surface.
    Meanwhile, it is worth mentioning that the sDBI possesses strong physical motivation, see \citep{ Alishahiha:2004eh,Chimento2010, BORDEMANN1993,BORDEMANN1994,Ogawa2000}.

	For the Lagrangian given in Eq.~(\ref{equ:lag}), applying the principle of least action gives the Einstein field equation:
	\begin{eqnarray}
		R_{\mu\nu } - \frac{1}{2}R g_{\mu\nu } =  -\kappa \left( T_{\mu\nu}^{(\Lambda_I)} + T_{\mu\nu}^{(\Lambda_{II})} + T_{\mu\nu}^{(m)}\right), \label{Einstein_equ}
	\end{eqnarray}
	where $R_{\mu\nu}$ is the Ricci tensor,  $T_{\mu\nu}^{(\Lambda_I)} =  - \Lambda_{I} g_{\mu\nu} $ and 
	\begin{eqnarray}
		T_{\mu\nu}^{(\Lambda_{II})} = \Lambda_{II}\left(\frac{\partial_\mu \phi \partial_\nu \phi }{\sqrt{1+\partial_\rho\phi\partial^\rho\phi}} 
		- g_{\mu\nu} \sqrt{1+\partial_\rho\phi\partial^\rho\phi}
		\right) \  \label{Energy-stress-general-form}
	\end{eqnarray}
	represent the energy-stress tensor of dark energy and dark matter in this model, respectively. Now our focus turns to the sDBI field. 
 According to Eq.~(\ref{Energy-stress-general-form}), in the flat, homogeneous, and isotropic background of the Universe,  sDBI field can be treated as a perfect fluid characterized by the Equation of State (EoS, see Appendix \ref{app1})
	\begin{eqnarray}
		w  = - \frac{\Lambda_{II}^2}{\rho^2}\ ,\label{equ:eos1}
	\end{eqnarray}
	where $w \equiv P / \rho$, $P$ and $\rho$ denoting the pressure and mass density of the sDBI field, respectively. 
    The evolution of $\rho$ and $w$ regard to scale factor $a$ can be derived as (see Appendix \ref{app1})
    \begin{eqnarray}
        \rho(a) = \Lambda_{II}a_d^3 \sqrt{a_d^{-6}+a^{-6}} \equiv \rho_s \sqrt{a_d^{-6}+a^{-6}}
    \end{eqnarray}
    and 
    \begin{eqnarray}
        w(a) = - \frac{1}{1 + \left( \frac{a_d}{a} \right)^6 }\ ,
    \end{eqnarray}
    respectively.
    Here, the scale factor  $a$ is normalized to unity at the present time, and $a_d$ is a free parameter that we call as decay parameter. 
    
	Moreover, considering a linear perturbation in the homogeneous Universe, the sound speed of the sDBI field can be given by (see Appendix \ref{app2})
	\begin{eqnarray}
		c_s^2 = c_a^2 = - w \ , \label{equ:ss}
	\end{eqnarray}
	where $c_s\equiv \delta P / \delta \rho$ and $c_a\equiv dP/d\rho $ are the effective and adiabatic sound speed, respectively. The EoS and sound speed provide sufficient information to complete the scalar linear evolution equations of Universe \citep{Hu_1998, hu2004covariant}.	
	
\begin{figure}
		\centering
		\subfigure[]{
			\includegraphics[width=0.9\linewidth]{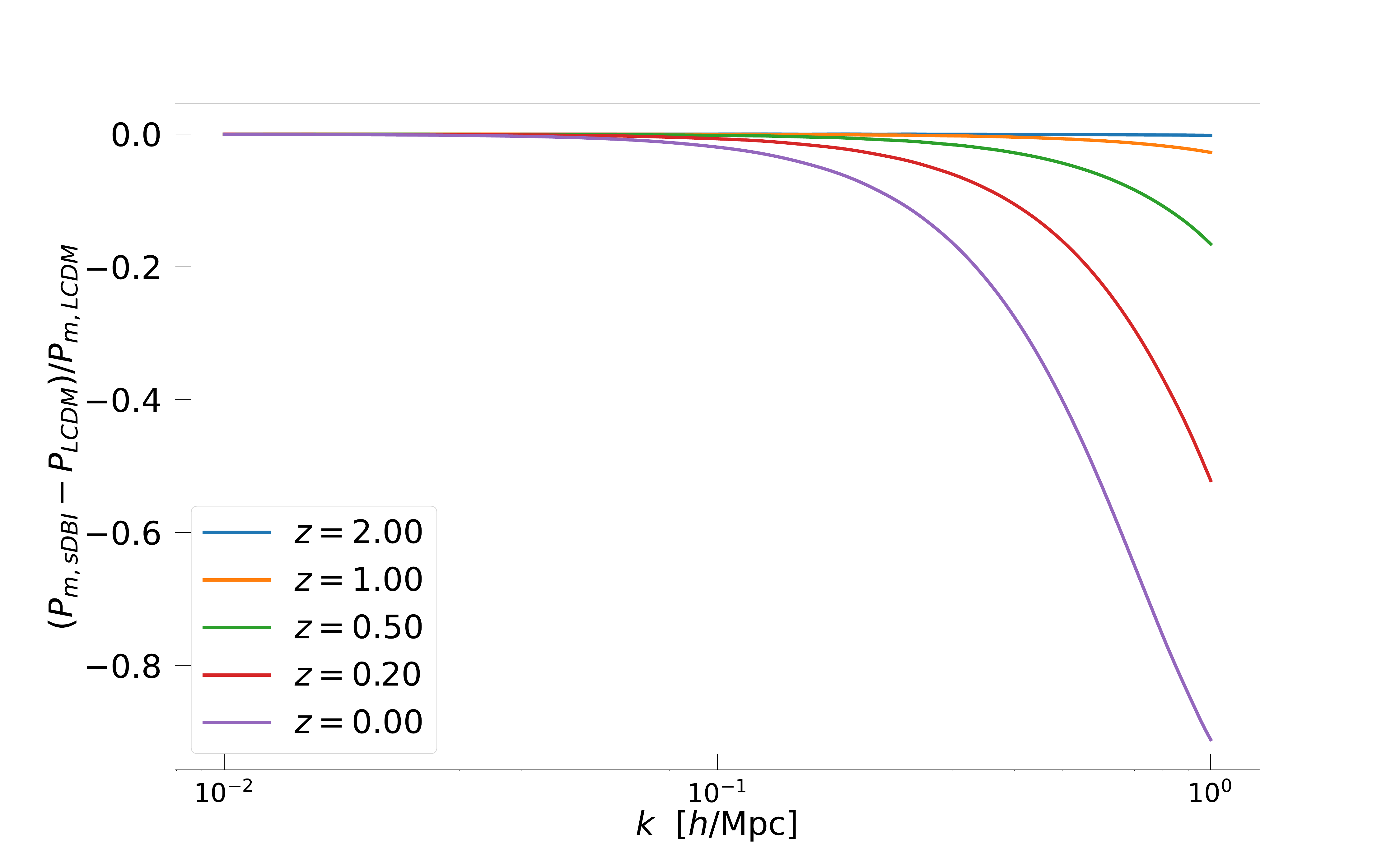} 
			\label{fig:dpk_vs_k}    }
		\subfigure[ ] {
			\includegraphics[width=0.9\linewidth]{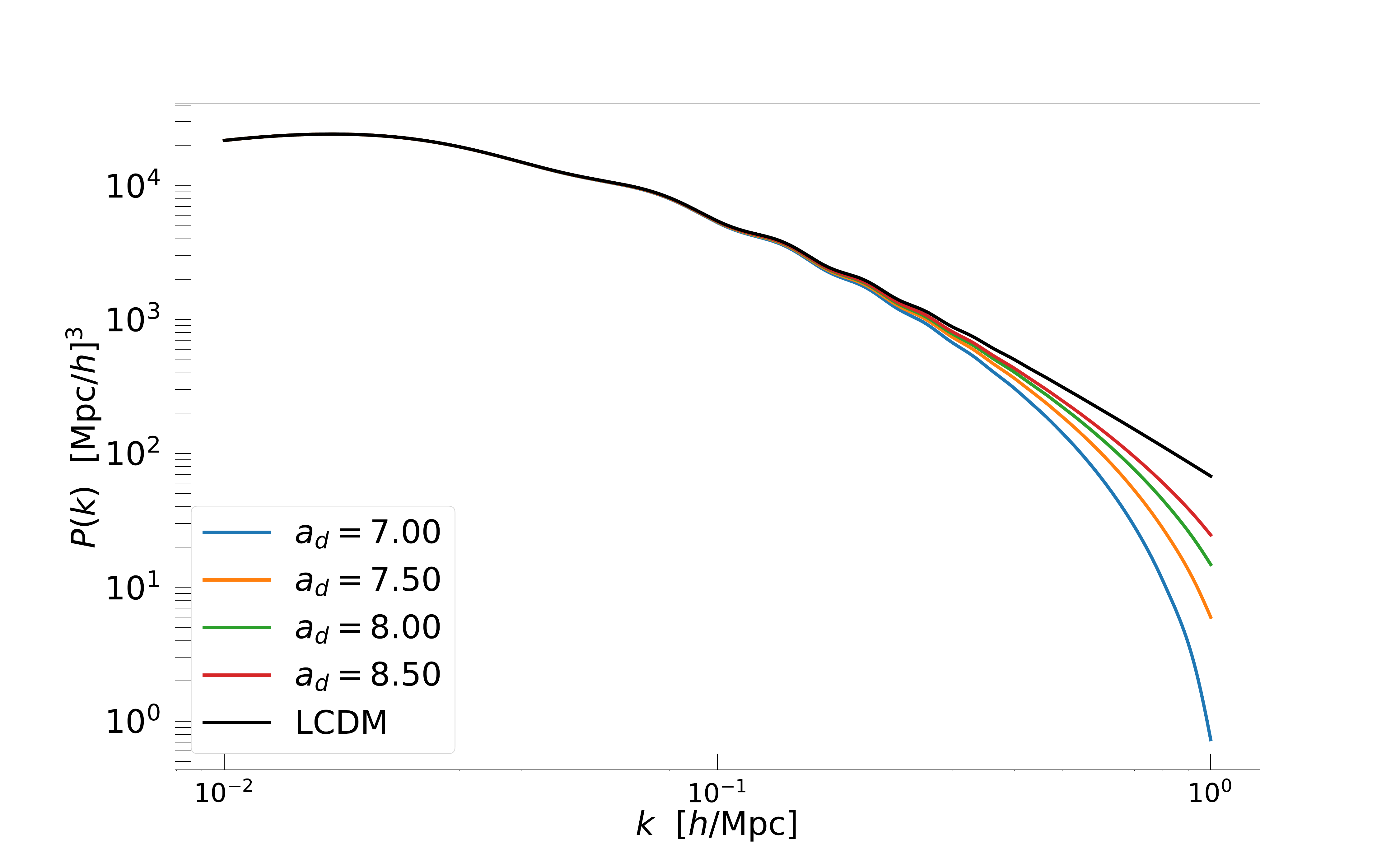} 
			\label{fig:pk_vs_k}
		}
		\caption{(a) The suppression of sDBI on matter power spectrum for different redshifts, where $\Delta P_k \equiv P_k^{(\mathrm{sDBI})} - P_k^{(\Lambda \mathrm{CDM})}$ with $a_d$ set to $3.8$. 
			(b) The matter power spectrum at redshift $z=0$ for different $a_d$s. When $a_d\to \infty$, the power spectrum asymptotically approaches the $\Lambda$CDM model (black line). 
			In both figures, the fixed parameters are taken from the best fit of the sDBI model in Table~\ref{tab:bestfit_mean}. }
		
		\label{fig:pk_k}
	\end{figure}

	The dark matter with the above form EoS and sound speed has such properties that during the early stages ($a\ll a_d$), it behaves similarly to the pressure-less standard cold dark matter, but at the late stages ($a$ close to $a_d$), it exhibits a certain sound speed and pressure, which leads to the smoothing out the structures that formed during the early stages \citep{PhysRevD.67.063509,BILIC200217,salahedin2020,Adil2023}.
    This may provide an explanation for the smoother Universe at low-redshift.
    In Fig.~\ref{fig:dpk_vs_k}, we illustrate the difference between the linear matter spectra of the sDBI and the $\Lambda$CDM model. 
    It is evident that the sDBI suppresses the power spectra in the late-stage Universe. The value of the decay parameter will greatly influence this process. Fig.~\ref{fig:pk_vs_k} shows the power spectra for different $a_d$s at $z=0$. As $a_d$ tends towards infinity, the sDBI model will degenerate to $\Lambda$CDM.
    Some similar late-time schemes have also been proposed \citep{poulin2022sigma, 2023arXiv230816183L}, offering viable solutions to the $S_8$ tension. The sDBI may serve as a theoretical framework for these late-time schemes.
	

    Note that in the non-linear region,  we strictly need to consider Eq.~(\ref{Einstein_equ}) and the evolution equation for $\phi$
	\begin{eqnarray}
		\left(\frac{1}{2} \partial_\mu \log\left( -g \right) + \partial_\mu \right) \frac{\partial^\mu \phi }{\sqrt{1+\partial_\nu\phi \partial^\nu \phi}}=0 \ , \label{minimal_serface_equ}
	\end{eqnarray}
	which represents a general minimal surface equation. However, in this work, we focus solely on the scalar linear perturbation, as it dominates the evolution of dark matter, especially on large scales and during the early stages of our Universe.
	
	\textit{Constraints by the observations.} --To demonstrate that the sDBI model can alleviate the $S_8$ tension, we perform a series of constraints using different observational data sets. We begin with the baseline of \textit{Planck2018}, which combines the $TT$, $TE$, $EE$ and low-$E$ angular power spectra of the CMB to constrain the cosmological parameters\cite{Planck2018}. This base line analysis is advantageous as it avoids model-dependent non-linear effects that may introduce uncertainties \citep{smith2003, Mead2016}. For the low-redshift probes, we employ the WL shear catalog from KiDS1000\citep{Kuijken2019, Asgari2021} and the GC data from SDSS-III BOSS\citep{Alam2017}. In these analyses, we treat the high-redshift probe (CMB) and low-redshift probes (WL and GC) separately, rather than combining them. If the two data sets can give a consistent result, it will provide stronger evidence in support of the model's validity. In addition, we simultaneously constrain the $\Lambda$CDM model as a control group for comparison purposes.

	Utilizing the EoS (\ref{equ:eos1}) and sound speed (\ref{equ:ss}), we have adapted the background and perturbation equations, as implemented in the Boltzmann code \texttt{CLASS-v3.2.0}\citep{Diego_Blas_2011} \footnote{\url{https://lesgourg.github.io/class_public/class.html}}, to calculate the linear evolution of the Universe. And a public Markov Chain Monte Carlo (MCMC) sampler \texttt{M\scriptsize ONTE\normalsize P\scriptsize YTHON\normalsize-v3.2}\citep{Thejs2019, Audren_2013}\footnote{\url{https://baudren.github.io/montepython.html}} was used. We perform all the MCMC samplings for our constraints using the Metropolis-Hastings algorithm implemented in \texttt{M\scriptsize ONTE\normalsize P\scriptsize YTHON\normalsize-v3.2}.
	
	To constrain this model with the \textit{Planck2018} baseline, we adopt a flat prior on certain nuisance parameters in the \textit{Planck} likelihood (\texttt{Plik}) \citep{Planck2018likelihood}, as well as on the cosmological parameters $\left\{\omega_b, \Omega_s, h, A_s, n_s, \tau_{reio}, a_d\right\}$, where 
	\begin{eqnarray}
		\Omega_s \equiv \frac{ \rho_s}{\rho_{cr}} \equiv \frac{8\pi G }{3H_0^2} \rho_s \
	\end{eqnarray}
	is the reduced dark matter density in our model. The names and priors of the base cosmological parameters are provided in Table~\ref{tab:prior}.  For comparison, we have also conducted a parallel $\Lambda$CDM constraint. Note that in all the analyses, we assume a flat universe with zero spatial curvature ($\Omega_K=0$). Additionally, our neutrino model follows the same configuration as \textit{Planck2018}, consisting of two massless species and one massive species with a mass of $0.06$ eV.
	
	\begin{table}
		\caption{The Symbols and Prior of cosmological parameters, where the infinity symbols $\pm \infty$ mean that we don't limit the upper/lower bound of the corresponding parameter.}
		\begin{ruledtabular} 
			\begin{tabular}{ccc}
				\textrm{Parameter }& \textrm{Symbol}&\textrm{Prior} \\ 
				\hline 
				Baryon density             & $\omega_b$                             & [$-\infty$, $\infty$] \\
				Dark matter density    &$\Omega_s$                             & [0,       $\infty$] \\
				Inverse of Decay parameter          & $a_d^{-1}$                             & ($0,\infty$] \\
				Reduced Hubble constant          & $h$                                & [$-\infty$, $\infty$] \\
				Scalar fluctuation amplitude& $A_s$                 & [$-\infty$, $\infty$] \\ 
				Scalar spectral index & $n_s$                               & [$-\infty$, $\infty$] \\
				Reionization optical depth & $\tau_{reio}$                  &[0.004, $\infty$] \\ 
			\end{tabular}   \label{tab:prior}
		\end{ruledtabular}
	\end{table}

	The posterior distributions obtained with \textit{Planck2018} baseline are presented in Table~\ref{tab:bestfit_mean}.  The Markov chain employed in the analysis satisfies the Gelman-Rubin convergence criterion with $R-1 \approx 10^{-3}$, indicating robust convergence. 
    Furthermore, our constraints on the $\Lambda$CDM model are consistent with the results reported by the \textit{Planck2018} collaboration \citep{Planck2018}, validating the accuracy of this analysis.
	
	The results reveal slight differences in common cosmological parameters between the sDBI and $\Lambda$CDM. However,  significant discrepancies have been observed in the structure growth parameter $S_8$.The sDBI model yields values of $(\Omega_m, S_8) =( 0.3199_{-0.0095}^{+0.0095}, 0.7448_{-0.21}^{+0.031}) $ , which agree with the results from \textit{K1K-3$\times$2pt} within $1\sigma$ and clearly deviate from the result given by \textit{Planck2018}.
	
	\begin{table}
		\renewcommand{\arraystretch}{1.5}
		\caption{The best-fit values,  mean values, 68\% credible intervals and $\chi^2$ for the sDBI and $\Lambda$CDM models from \textit{Plank} CMB power spectrum. The first 8 parameters consist of the base parameters, which include 5 common parameters and 3 peculiar parameters specific to each model ($\Omega_s$ and $a_d$ for sDBI, and $\Omega_{cdm}$ for $\Lambda$CDM). The last 4 parameters are derived quantities, where $\Omega_{vac}$ is defined as $\Omega_{\Lambda}$ and $\Omega_{\Lambda_I}$ in the $\Lambda$CDM and sDBI models, respectively.  Additionally, we use the notation $\Omega_m \equiv 1 - \Omega_{vac}$ for both models. This term represents the sum of the density parameters of all components in the Universe, excluding the contribution from dark energy.  
  }
		\begin{ruledtabular}
			\begin{tabular}{ccccc} 
				&  \multicolumn{2}{c}{sDBI}                   & \multicolumn{2}{c}{$\Lambda$CDM}          \\
				Parameters              &best fit & mean$^{+\sigma}_{-\sigma}$         & best fit  & mean$^{+\sigma}_{-\sigma}$     \\
				\hline 
				$\Omega{}_{s }$         &$0.2687$ & $0.2688_{-0.0089}^{+0.0087}$      & -         & -                              \\
				$a^{-1}_{d }$                &$0.1476$ & $0.1190_{-0.078}^{+0.080}$           & -         & -                              \\
				$\Omega_{cdm}$          &  -      &  -                                  &$0.2686$ & $0.2678_{-0.0075}^{+0.0086}$  \\
				\hline 
				$100~\omega{}_{b }$    &$2.233$ & $2.233_{-0.015}^{+0.014}$          &$2.224$ & $2.233_{-0.015}^{+0.015}$      \\  
				$h$                    &$0.6709$ & $0.6706_{-0.0067}^{+0.0068}$       &$0.6704$ & $0.6713_{-0.0066}^{+0.0060}$   \\  
				$10^{9}A_{s }$        &$2.089$ & $2.110_{-0.031}^{+0.037}$         &$2.089$ & $2.107_{-0.033}^{+0.033}$      \\ 
				$n_{s }$                &$0.9628$ & $0.9619_{-0.0044}^{+0.0046}$      &$0.9630$ & $0.9628_{-0.0040}^{+0.0042}$   \\
				$\tau{}_{reio }$       &$0.05017$ & $0.05524_{-0.0069}^{+0.0078}$       &$0.05101$ & $0.05472_{-0.0072}^{+0.0079}$   \\
				\hline 
				$\Omega{}_{vac }$       &$0.6802$ & $0.6800_{-0.0095}^{+0.0095}$          &$0.6804$ & $0.6811_{-0.0095}^{+0.0080}$     \\  
				$S_{8 }$                 &$0.7751$ & $0.7448_{-0.21}^{+0.031}$          &$0.8364$ & $0.8384_{-0.016}^{+0.017}$     \\  
				$\sigma_8$              &$0.7508$ & $0.7214_{-0.21}^{+0.019}$         &$0.8105$ & $0.8133_{-0.0077}^{+0.0083}$    \\
				$\Omega{}_{m }$         &$0.3197$ & $0.3199_{-0.0095}^{+0.0095}$        &$0.3195$ & $0.3188_{-0.0080}^{+0.0095}$     \\     
                $\chi^2 $ &  \multicolumn{2}{c}{2749.20 } & \multicolumn{2}{c}{2749.38 } \\
			\end{tabular} \label{tab:bestfit_mean}
		\end{ruledtabular}
	\end{table}
    
    Note there is a relatively larger credible interval for both $S_8$ or $\sigma_8$. To further understand it, we fix the cosmological parameters except $a_d$ to the best fit in Table~\ref{tab:bestfit_mean}, then calculate $S_8$ and $\chi_\nu^2$ for different $a_d$s. Here the reduced chi-square $\chi^2_{\nu}$ is defined as chi-squared divided by the degrees of freedom
	\begin{eqnarray}
		\chi^2_\nu  \equiv \frac{1}{N-n} \chi^2 \equiv \frac{1}{N-n} \sum_{i=1}^{N} \frac{\left(O_i - C_i\right)^2}{\sigma_i^2}
	\end{eqnarray} 
	with $O_i$ and $\sigma_i$ the $i$-th observed mean value and measure error, respectively, $C_i $ the corresponding prediction, $N$ the number of observed values, $n$ the number of fitted parameters.
    \begin{figure}
        \centering
        \includegraphics[width=0.9\linewidth]{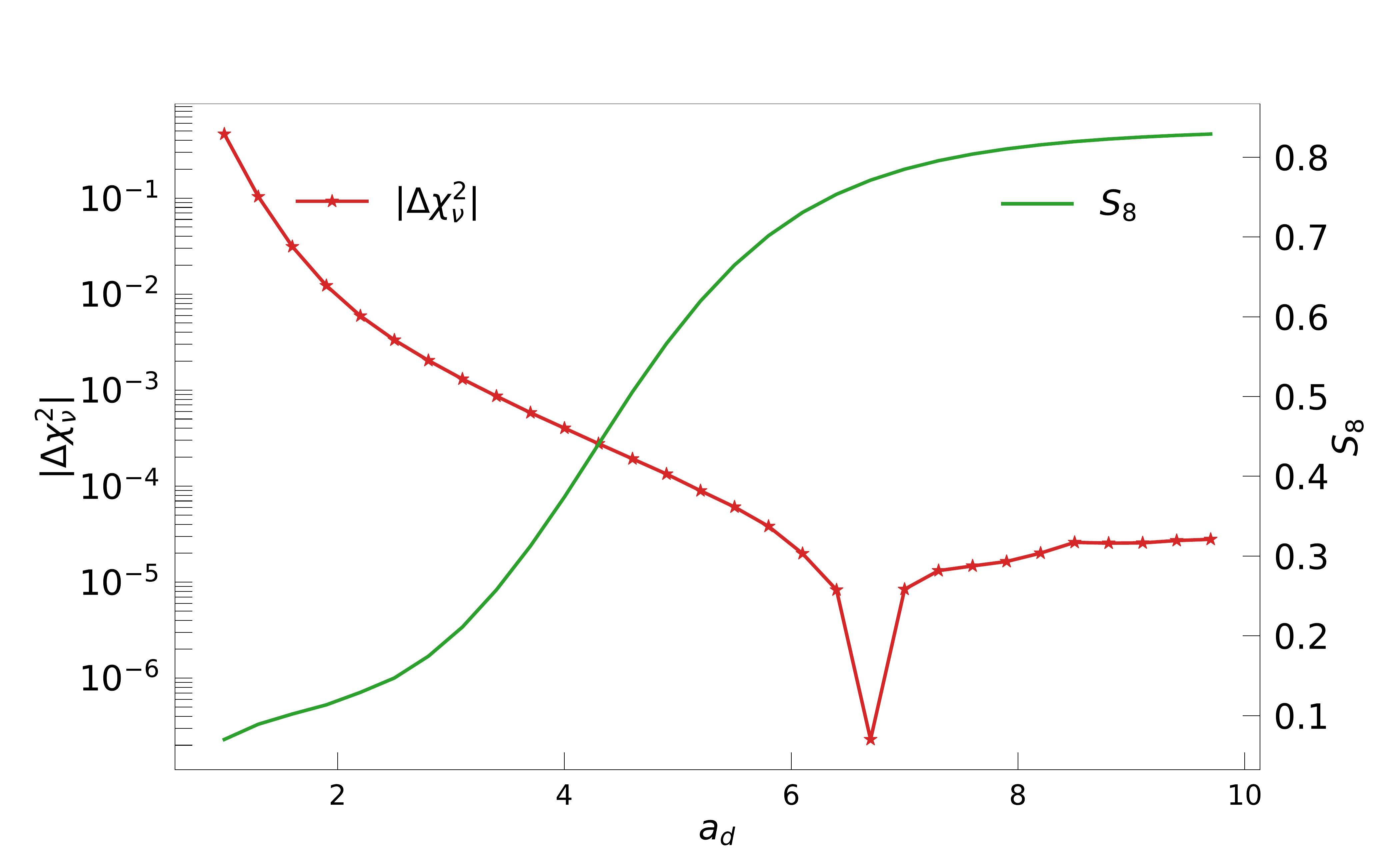}
        \caption{The $|\Delta\chi_\nu^2|$ (red) and $S_8$ (green) regard to the decay parameter $a_d$, where $\Delta \chi_\nu^2 \equiv \chi_\nu^2 - \chi^2_{\nu,b}$ with $\chi^2_{\nu,b} \approx 1.038422 $ the chi-square for the best fit. The drop near $a_d=6.5$ is due to its proximity to the best fit.  }
        \label{fig:chi2}
    \end{figure}

    The results are shown in Fig.~\ref{fig:chi2}, where $\Delta \chi_\nu^2 \equiv \chi_\nu^2 - \chi^2_{\nu,b}$ with $\chi^2_{\nu,b} \approx 1.038422 $ the chi-square for the best fit. Visibly, the fitting is almost as good as the best fit if $a_d\gtrsim 6$, where,  however, the $S_8$ can still vary from about 6.5 to about 0.8. In other words, the sDBI model can decrease the value of $S_8$ without influence CMB power spectra.
     In addition, it is worth noting that based on this mechanism,  the sDBI model does not aggravate the Hubble tension.

	\begin{figure*}
		\centering
		\subfigure[]
		{
			\includegraphics[width=0.45\linewidth]{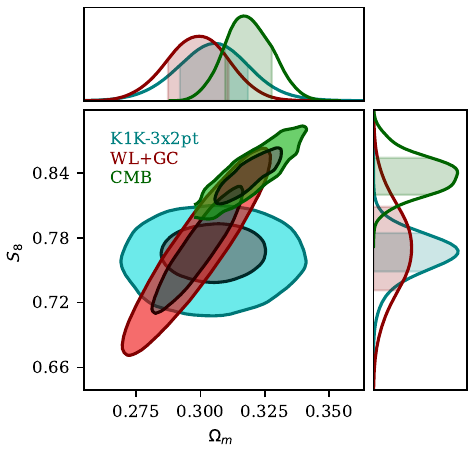}\label{fig:S8_vs_Om_lcdm}
		}
		\subfigure[]
		{
			\includegraphics[width=0.45\linewidth]{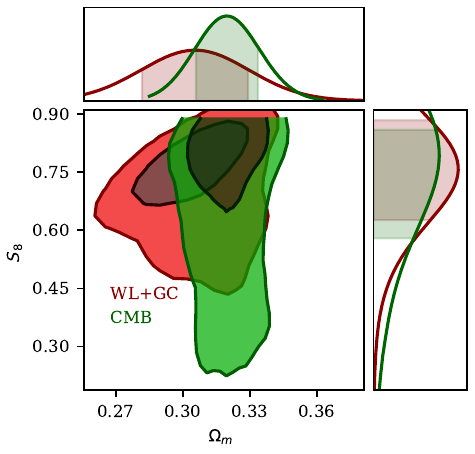}\label{fig:S8_vs_Om_sdbi}
		}
		\caption{ The posterior distributions given by high-redshift probe (CMB) and low-redshift probes (WL+GC). For low-redshift probes, the five common parameters are fixed to the values given in Table~\ref{tab:bestfit_mean} for respective models. (a) The posterior distributions of $\Omega_m$ and $S_8$ for $\Lambda$CDM model, where we also plot the result from \textit{K1K-3$\times$2pt} for comparison. Note that even cut off the non-linear effect, our low-redshift constraint is consistent with \textit{K1K-3$\times$2pt}, and the tension of $S_8$ still exists. (b) Same as (a) but for the sDBI model. No tension exists. }
	\end{figure*}

	
    After constraining the models using \textit{Planck2018} CMB power spectra, we proceed to perform a combined constraint by incorporating low-redshift probes, specifically WL and GC. Once again, we carry out parallel constraints for both the sDBI and $\Lambda$CDM models.
    Our approach to the correlation function involves mitigating non-linear effects due to their unavailability in our model. For WL, we employ the correlation function $\xi_+(\theta)$ and implement a truncation to the small-scale region ($\theta<10$) using the KiDS cosmology analysis pipeline \texttt{kcap}\citep{Joachimi2021,ZUNTZ201545}.
    The truncation is achieved through the following steps: First, we separately calculate the correlation function data vectors with and without the non-linear effect with \texttt{kcap}, and label them by $\mathbf{\xi}_+^{NL}$ and $\mathbf{\xi}_+^{L}$, respectively. Subsequently, we calculate the relative distance between the output data vectors, denoted as $d_c$ and defined by $||\mathbf{\xi}_+^{NL} - \mathbf{\xi}_+^{L}|| / ||\mathbf{\xi}_+^{NL}||$ with $||\cdot || \equiv \sqrt{\langle \cdot, \cdot\rangle}$. If $d_c$ exceeds a threshold of $10^{-2}$, we utilize \texttt{kcap} to remove some small-scale data points. This process is iteratively repeated until $d_c$ falls below $10^{-2}$.
	Note that we have excluded the correlation function $\xi_-$ from our analysis as the non-linear effects on $\xi_-$ are difficult to mitigate.
 
	For GC, our analysis specifically focuses on the measurements of the baryon acoustic oscillations (BAO) while excluding the consideration of redshift-space distortions. 
	Due to the strict elimination of the non-linear effect, the constraint on the five common base parameters becomes weaker.
        Hence, for both the sDBI and $\Lambda$CDM models, we fix these parameters according to their respective best-fit values in Table~\ref{tab:bestfit_mean}. 

	The $\Lambda$CDM model constraint yields $(\Omega_m, S_8) = (0.299_{-0.0105}^{+0.011}, 0.770_{-0.035}^{+0.0371})$, which is consistent with the results obtained from \textit{K1K-3$\times$2pt}. However, it is evident that $S_8$ remains lower compared to the \textit{Planck2018} baseline. As depicted in Fig.~\ref{fig:S8_vs_Om_lcdm}, the tension between low-redshift probes and the CMB persists.
	
	In contrast, for the sDBI model, the tension in $S_8$ is absent. As illustrated in Fig.~\ref{fig:S8_vs_Om_sdbi}, the constraint provides $(\Omega_m, S_8) = (0.306_{-0.014}^{+0.014},0.7426_{-0.085}^{+0.054}) $, which is consistent within 1$\sigma$ compared to our constraint using the \textit{Planck2018} baseline. 

    \textit{Conclusion and Discussion.} --
    In this article, we propose a so-called surface-type Dirac-Born-Infeld (sDBI) field as a dark matter candidate to relieve the $S_8$ tension. We have conducted a parallel investigation of t9ihe background and linear perturbation evolution for both the sDBI and standard $\Lambda$CDM models. The MCMC analysis with the data from early Universe (CMB) and late Universe (WL) reveals that the $S_8$ tension persists in the $\Lambda$CDM model even when considering only linear perturbations. 
    This suggests that modifying the non-linear model such as \texttt{halofit}\citep{smith2003} or \texttt{hmcode}\citep{Mead2016}, is unlikely to resolve the tension effectively. On the other hand, the sDBI model, within the scope of the data sets we have considered, successfully alleviates the $S_8$ tension. 

    To obtain more accurate constraints on the parameter $a_d$, it is necessary to consider non-linear effects, which can be studied by N-body simulations. In the non-relativistic approximation,
    for the fluid described by Eq.~(\ref{equ:eos1}), we can introduce an effective potential (see Appendix \ref{app3})
	\begin{eqnarray}
		h \equiv  - \int_\rho^\infty \frac{dP(\rho')}{\rho'} = - \frac{1}{2} \frac{\Lambda_{II}^2}{\rho^2} \label{def:h}
	\end{eqnarray}
	to substitute the effect of pressure.  Using an effective potential to describe the sDBI has the advantage that it can be easily incorporated into N-body simulation codes.
     
     The potential described in Eq.~(\ref{def:h}) acts as a contrary effect from gravity, and the gradient of the potential increases in conjunction with the decrease in energy density $\rho$ over time. The increasing external force can partially disrupt the structures that formed in the early universe, potentially leading to the formation of galaxies with a lack of dark matter \citep{Dokkum2018, Ogiya2022}. Furthermore, the external force may completely destroy certain dwarf galaxies \citep{Kase2007}. We also anticipate a reduction in redshift-space distortion over time, as the relative motion between dark matter halos decelerates. The $S_8(z)$ test could serve as a robust assessment for this model, as the external potential is expected to significantly diminish the fluctuation amplitude of matter in the later stages. Meanwhile, there is a double coincidence problem inherent in $a_d$ and $\Omega_{II}$. In order to address this issue, a more fundamental theory will be required in the future to provide a resolution. Given the intricacy of these topics, we defer their exploration to future work.

    
	\begin{acknowledgments}
		Xingpao Suo and Xi Kang acknowledge the support from the National Key Research and Development Program of China (No.2022YFA1602903), the NSFC \ (No. 11825303, 11861131006), the science research grants from the China Manned Space project with No. CMS-CSST-2021-A03, CMS-CSST-2021-A04, the Fundamental Research Funds for the Central Universities of China \ (226-202  2-00216) and the start-up funding of Zhejiang University.	Huanyuan Shan acknowledges the support from NSFC of China under grant 11973070, Key Research Program of Frontier Sciences, CAS, Grant No. ZDBS-LY-7013, Program of Shanghai Academic/Technology Research Leader, and the science research grants from the China Manned Space Project with NO. CMS-CSST-2021-A01, CMS-CSST-2021-A04. We thank Joe Zuntz, Benjamin St\"olzner, Sunny Vagnozzi and Yanhong Yao for helpful discussions. Moreover, we thank the anonymous referees for improving the quality of our manuscript. 
	\end{acknowledgments}
    \appendix

    \section{The Evolution and Fluid Equivalence of sDBI Field}
    \subsection{The Background Evolution}\label{app1}
    For the curvature-free, homogeneous, and isotropic background of the Universe, the spacetime is described by the flat Friedmann-Lemaître-Robertson-Walker (FLRW) metric
    \begin{eqnarray}
        ds^2
        &  = a^2(\tau)\left( - d\tau^2 + \delta_{ij}dx^i dx^j \right) \ ,\label{equ:bakgrd}
    \end{eqnarray}
    where $a$ is the scale factor, $\tau\equiv \int dt/a$ with cosmic time $t$ is conformal time,  $(x^1,x^2,x^3)$ are the coordinates of space, $\delta_{ij}$ is Kronecker symbol. 
    In such a Universe, the background evolution of the sDBI field is also solely dependent on time. Consequently, the energy-stress tensor (\ref{Energy-stress-general-form}) can be simplified as follows:  
    \begin{eqnarray}
        &T_{00} =  g_{00} \Lambda_{II}\frac{1}{\sqrt{1-\phi'^2/a^2}} \equiv g_{00} \rho \label{equ:Tmn1} \\
        &T_{ii} = -g_{ii} \Lambda_{II}\sqrt{1-\phi'^2/a^2} \equiv  g_{ii}P \label{equ:Tmn2} \\ 
        &T_{ij} = T_{0i} = 0 \label{equ:Tmn3}
    \end{eqnarray}
    where $\rho$ and $P$ are the energy density and pressure of the sDBI field, respectively, $i,j=1,2,3$ with $i\neq j$, $g_{\mu\nu}$ is the metric tensor, $'$ denotes a derivative with respect to conformal time. Indeed, in this scenario, it follows that $P\rho = -\Lambda_{II}^2$, so sDBI field is equivalent to an ideal fluid with the EoS
    \begin{eqnarray}
        w \equiv \frac{P}{\rho} = -\frac{\Lambda_{II}^2}{\rho^2}\ . \label{equ:eos2}
    \end{eqnarray}
    By utilizing the Einstein field equation (\ref{Einstein_equ}), one can derive the conservation law
    \begin{eqnarray}
        T^{\mu}_{\nu;\mu} = G^{\mu}_{\nu;\mu} = 0\ .\label{equ:conser}
    \end{eqnarray}
    For the background of the Universe, Eq.~(\ref{equ:conser}) can be simplified to 
    \begin{eqnarray}
        \rho' + 3 \mathcal H  \left(\rho + P\right) = 0 \label{equ:conser1}
    \end{eqnarray}
    where $\mathcal H \equiv a'/a$ is the conformal Hubble parameter. One can change the independent variable in Eq.~(\ref{equ:conser1}) from time $t$ to scale factor $a$
    
    \begin{eqnarray}
        \frac{d\rho}{d\log a} + 3 (\rho + P) = 0\ . \label{equ:conservation}
    \end{eqnarray}
    Combining Eq.~ (\ref{equ:conservation}) and (\ref{equ:eos2}) one gets the evolution of the energy density regard to the scale factor
    \begin{eqnarray}
        \rho(a) = \Lambda_{II}a_d^3 \sqrt{a_d^{-6}+a^{-6}} \equiv \rho_s \sqrt{a_d^{-6}+a^{-6}}
    \end{eqnarray}
    and 
    \begin{eqnarray}
        w(a) = - \frac{1}{1 + \left( \frac{a_d}{a} \right)^6 }\ ,
    \end{eqnarray}
    where the integration constant $a_d$ is introduced, which we refer to as the decay parameter.

    \subsection{The Linear Perturbation Evolution}\label{app2}
    When considering linear perturbations on the background metric in Eq.~(\ref{equ:bakgrd}) and neglecting vector and tensor perturbations while adopting the Newtonian gauge, one can express the perturbed metric as
    \begin{align}
        ds^2 = a^2(\tau)\left( -(1+2\Psi) d\tau^2 + (1+2\Phi)\delta_{ij} dx^i dx^j \right)\ ,\label{equ:pertb}
    \end{align}
    where $\Psi $ and $\Phi$ are two spatial scalars referred to as Newtonian and curvature potential, respectively.
    By substituting Eq.~(\ref{equ:pertb}) into the Einstein equation (\ref{Einstein_equ}) and considering only terms up to linear order, we can obtain the scalar perturbation equations for the sDBI field
    \begin{align}
        \delta' = - 3\left( c_s^2 - w \right)\mathcal H  \delta - (1 + w) (\theta + 3 \Phi') \\ \nonumber 
        \theta' = -\left( 1- 3c_s^2 \right) \mathcal H \theta - \nabla^2 \left( \frac{c_s^2}{1+w} \delta + \Psi \right) 
    \end{align}
    where $\delta \equiv (\rho-\bar\rho) / \bar\rho $ is the density contrast with the density $\rho$ and averaged density $\bar\rho $, $ \theta \equiv \nabla \cdot \mathbf u$ is the velocity divergence with $\mathbf u$ the velocity of the fluid elementary, $c_s^2 \equiv \delta P / \delta \rho$ with the pressure perturbation $\delta P $ and density perturbation $\delta \rho $ is the effective sound speed.

    In realistic calculations, the gravitational potentials $\Psi$ and $\Phi$ are determined by considering contributions from all components of the Universe. These contributions can be obtained through linearized Einstein equations. The conformal Hubble parameter $\mathcal{H}$ is obtained during the calculation of background evolution. Consequently, the effective sound speed $c_s^2$ becomes the only degree of freedom that requires specification.

    To obtain the effective sound speed $c_s^2$ in the sDBI model, one considers a perturbation $\delta\phi$ on the background sDBI field $\phi$. The perturbed energy-stress caused by $\delta \phi $ is given by
    \begin{align}
        \delta T^\mu_\nu  = \frac{\partial T^\mu_\nu }{\partial\phi } \delta\phi 
        + \frac{\partial T^\mu_\nu}{\partial (\partial_\alpha  \phi)  } \delta (\partial_\alpha  \phi) 
        + \frac{\partial T^\mu_\nu}{\partial g_{\alpha \beta } } \delta g_{\alpha \beta } \ , \label{equ:general_dT}
    \end{align}
    where $T^\mu_{\nu} $ is given by Eq.~(\ref{equ:Tmn1}-\ref{equ:Tmn3}).  The meticulous but straightforward calculation gives the linear order perturbation 
    \begin{align}
        &\delta \rho = -\delta T^0_0 =  \frac{\Lambda_{II}}{\left( 1-\phi'^2/a^2\right)^{3/2} a^2} \left( \frac{1}{2} \phi'^2 \delta g^{00} + \phi' \delta \phi' \right)\\
        &\delta P = \delta T_{i}^i = \frac{\Lambda_{II}}{\left( 1-\phi'^2/a^2\right)^{1/2} a^2} \left( \frac{1}{2} \phi'^2 \delta g^{00} + \phi' \delta \phi' \right)\ ,
    \end{align}
    which gives
    \begin{align}
        c_s^2 \equiv \frac{\delta P }{\delta \rho } = 1- \frac{\phi'^2 }{a^2 } = -w\ .
    \end{align}
    Note that the adiabatic sound speed of the sDBI field is 
    \begin{align}
        c_a^2 \equiv \frac{d P}{d\rho } = \frac{\Lambda_{II}^2}{\rho^2 } = -w \ .
    \end{align}
    It is evident that $c_s^2 = c_a^2$, and this relationship is not coincidental. The equation of state (EoS) given by Eq.~(\ref{equ:eos2}) demonstrates that the sDBI field can be regarded as a barotropic fluid, where the pressure $P$ solely depends on the mass density $\rho$. In the case of a barotropic fluid, the effective sound speed is equal to the adiabatic sound speed.

    \subsection{The Non-relativistic Fluid Equivalance}
    \label{app3}
     Assuming that Newton's laws still apply to the sDBI fluid, the momentum equation can be expressed as
    \begin{eqnarray}
        \frac{D\mathbf u}{Dt} =- \frac{\nabla P}{\rho} - \nabla \Psi\ ,\label{equ:Euler}
    \end{eqnarray}
    where $\mathbf u$ is the velocity of the fluid elementary, $P$ and $\rho$ are the pressure and mass density of the fluid, respectively, $\Psi$ is the Newtonian gravitational potential, $D/Dt\equiv \partial/\partial t + \mathbf u \cdot \nabla$ is the material derivative. Introducing an effective potential $h$, one can rewrite Eq.~(\ref{equ:Euler}) as
    \begin{eqnarray}
        \frac{D\mathbf u}{Dt} =- \nabla (h + \Psi) \ ,
    \end{eqnarray}
    where $h$ is defined by 
    \begin{align}
        h \equiv  - \int_\rho^\infty \frac{dP(\rho')}{\rho'} = - \frac{1}{2} \frac{\Lambda_{II}^2}{\rho^2}\ .
    \end{align}
    In the early stage of the Universe, $\rho \gg \Lambda_{II} $, thus $|h| \ll 1$ and the evolution is dominated by $\Psi $. However, in the late stage of the Universe, the role of the perturbation $h$ gradually becomes more significant. In regions with higher matter density, the gravitational potential tends to have a larger value. This behavior is contrary to gravity, where regions with higher matter density typically exhibit a smaller gravitational potential.
    
	\nocite{*}
	
	\bibliography{apssamp}

\begin{thebibliography}{49}%
\makeatletter
\providecommand \@ifxundefined [1]{%
 \@ifx{#1\undefined}
}%
\providecommand \@ifnum [1]{%
 \ifnum #1\expandafter \@firstoftwo
 \else \expandafter \@secondoftwo
 \fi
}%
\providecommand \@ifx [1]{%
 \ifx #1\expandafter \@firstoftwo
 \else \expandafter \@secondoftwo
 \fi
}%
\providecommand \natexlab [1]{#1}%
\providecommand \enquote  [1]{``#1''}%
\providecommand \bibnamefont  [1]{#1}%
\providecommand \bibfnamefont [1]{#1}%
\providecommand \citenamefont [1]{#1}%
\providecommand \href@noop [0]{\@secondoftwo}%
\providecommand \href [0]{\begingroup \@sanitize@url \@href}%
\providecommand \@href[1]{\@@startlink{#1}\@@href}%
\providecommand \@@href[1]{\endgroup#1\@@endlink}%
\providecommand \@sanitize@url [0]{\catcode `\\12\catcode `\$12\catcode
  `\&12\catcode `\#12\catcode `\^12\catcode `\_12\catcode `\%12\relax}%
\providecommand \@@startlink[1]{}%
\providecommand \@@endlink[0]{}%
\providecommand \url  [0]{\begingroup\@sanitize@url \@url }%
\providecommand \@url [1]{\endgroup\@href {#1}{\urlprefix }}%
\providecommand \urlprefix  [0]{URL }%
\providecommand \Eprint [0]{\href }%
\providecommand \doibase [0]{https://doi.org/}%
\providecommand \selectlanguage [0]{\@gobble}%
\providecommand \bibinfo  [0]{\@secondoftwo}%
\providecommand \bibfield  [0]{\@secondoftwo}%
\providecommand \translation [1]{[#1]}%
\providecommand \BibitemOpen [0]{}%
\providecommand \bibitemStop [0]{}%
\providecommand \bibitemNoStop [0]{.\EOS\space}%
\providecommand \EOS [0]{\spacefactor3000\relax}%
\providecommand \BibitemShut  [1]{\csname bibitem#1\endcsname}%
\let\auto@bib@innerbib\@empty
\bibitem [{\citenamefont {Abdalla}\ \emph {et~al.}(2022)\citenamefont {Abdalla}
  \emph {et~al.}}]{ABDALLA202249}%
  \BibitemOpen
  \bibfield  {author} {\bibinfo {author} {\bibfnamefont {E.}~\bibnamefont
  {Abdalla}} \emph {et~al.},\ }\href
  {https://doi.org/https://doi.org/10.1016/j.jheap.2022.04.002} {\bibfield
  {journal} {\bibinfo  {journal} {Journal of High Energy Astrophysics}\
  }\textbf {\bibinfo {volume} {34}},\ \bibinfo {pages} {49} (\bibinfo {year}
  {2022})}\BibitemShut {NoStop}%
\bibitem [{\citenamefont {{Di Valentino}}\ \emph {et~al.}(2021)\citenamefont
  {{Di Valentino}} \emph {et~al.}}]{DIVALENTINO2021102604}%
  \BibitemOpen
  \bibfield  {author} {\bibinfo {author} {\bibfnamefont {E.}~\bibnamefont {{Di
  Valentino}}} \emph {et~al.},\ }\href
  {https://doi.org/https://doi.org/10.1016/j.astropartphys.2021.102604}
  {\bibfield  {journal} {\bibinfo  {journal} {Astroparticle Physics}\ }\textbf
  {\bibinfo {volume} {131}},\ \bibinfo {pages} {102604} (\bibinfo {year}
  {2021})}\BibitemShut {NoStop}%
\bibitem [{\citenamefont {{Asgari, Marika}}\ \emph {et~al.}(2021)\citenamefont
  {{Asgari, Marika}} \emph {et~al.}}]{Asgari2021}%
  \BibitemOpen
  \bibfield  {author} {\bibinfo {author} {\bibnamefont {{Asgari, Marika}}}
  \emph {et~al.},\ }\href {https://doi.org/10.1051/0004-6361/202039070}
  {\bibfield  {journal} {\bibinfo  {journal} {A\&A}\ }\textbf {\bibinfo
  {volume} {645}},\ \bibinfo {pages} {A104} (\bibinfo {year}
  {2021})}\BibitemShut {NoStop}%
\bibitem [{\citenamefont {Abbott}\ \emph {et~al.}(2022)\citenamefont {Abbott}
  \emph {et~al.}}]{PhysRevD.105.023520}%
  \BibitemOpen
  \bibfield  {author} {\bibinfo {author} {\bibfnamefont {T.~M.~C.}\
  \bibnamefont {Abbott}} \emph {et~al.} (\bibinfo {collaboration} {DES
  Collaboration}),\ }\href {https://doi.org/10.1103/PhysRevD.105.023520}
  {\bibfield  {journal} {\bibinfo  {journal} {Phys. Rev. D}\ }\textbf {\bibinfo
  {volume} {105}},\ \bibinfo {pages} {023520} (\bibinfo {year}
  {2022})}\BibitemShut {NoStop}%
\bibitem [{\citenamefont {{Salvati, Laura}}\ \emph {et~al.}(2018)\citenamefont
  {{Salvati, Laura}} \emph {et~al.}}]{Salvati2018}%
  \BibitemOpen
  \bibfield  {author} {\bibinfo {author} {\bibnamefont {{Salvati, Laura}}}
  \emph {et~al.},\ }\href {https://doi.org/10.1051/0004-6361/201731990}
  {\bibfield  {journal} {\bibinfo  {journal} {A\&A}\ }\textbf {\bibinfo
  {volume} {614}},\ \bibinfo {pages} {A13} (\bibinfo {year}
  {2018})}\BibitemShut {NoStop}%
\bibitem [{\citenamefont {Ivanov}\ \emph {et~al.}(2020)\citenamefont {Ivanov},
  \citenamefont {Simonović},\ and\ \citenamefont {Zaldarriaga}}]{Ivanov_2020}%
  \BibitemOpen
  \bibfield  {author} {\bibinfo {author} {\bibfnamefont {M.~M.}\ \bibnamefont
  {Ivanov}}, \bibinfo {author} {\bibfnamefont {M.}~\bibnamefont {Simonović}},\
  and\ \bibinfo {author} {\bibfnamefont {M.}~\bibnamefont {Zaldarriaga}},\
  }\href {https://doi.org/10.1088/1475-7516/2020/05/042} {\bibfield  {journal}
  {\bibinfo  {journal} {Journal of Cosmology and Astroparticle Physics}\
  }\textbf {\bibinfo {volume} {2020}}\bibinfo  {number} { (05)},\ \bibinfo
  {pages} {042}}\BibitemShut {NoStop}%
\bibitem [{\citenamefont {Corasaniti}\ \emph {et~al.}(2021)\citenamefont
  {Corasaniti}, \citenamefont {Sereno},\ and\ \citenamefont
  {Ettori}}]{Corasaniti_2021}%
  \BibitemOpen
\bibfield  {number} {  }\bibfield  {author} {\bibinfo {author} {\bibfnamefont
  {P.-S.}\ \bibnamefont {Corasaniti}}, \bibinfo {author} {\bibfnamefont
  {M.}~\bibnamefont {Sereno}},\ and\ \bibinfo {author} {\bibfnamefont
  {S.}~\bibnamefont {Ettori}},\ }\href
  {https://doi.org/10.3847/1538-4357/abe9a4} {\bibfield  {journal} {\bibinfo
  {journal} {The Astrophysical Journal}\ }\textbf {\bibinfo {volume} {911}},\
  \bibinfo {pages} {82} (\bibinfo {year} {2021})}\BibitemShut {NoStop}%
\bibitem [{\citenamefont {{Heymans, Catherine}}\ \emph
  {et~al.}(2021)\citenamefont {{Heymans, Catherine}} \emph
  {et~al.}}]{Heymans2021}%
  \BibitemOpen
  \bibfield  {author} {\bibinfo {author} {\bibnamefont {{Heymans, Catherine}}}
  \emph {et~al.},\ }\href {https://doi.org/10.1051/0004-6361/202039063}
  {\bibfield  {journal} {\bibinfo  {journal} {A\&A}\ }\textbf {\bibinfo
  {volume} {646}},\ \bibinfo {pages} {A140} (\bibinfo {year}
  {2021})}\BibitemShut {NoStop}%
\bibitem [{\citenamefont {{Planck
  Collaboration}}(2020{\natexlab{a}})}]{Planck2018}%
  \BibitemOpen
  \bibfield  {author} {\bibinfo {author} {\bibnamefont {{Planck
  Collaboration}}},\ }\href {https://doi.org/10.1051/0004-6361/201833910}
  {\bibfield  {journal} {\bibinfo  {journal} {A\&A}\ }\textbf {\bibinfo
  {volume} {641}},\ \bibinfo {pages} {A6} (\bibinfo {year}
  {2020}{\natexlab{a}})}\BibitemShut {NoStop}%
\bibitem [{\citenamefont {Troxel}\ \emph {et~al.}(2018)\citenamefont {Troxel}
  \emph {et~al.}}]{DES2018}%
  \BibitemOpen
  \bibfield  {author} {\bibinfo {author} {\bibfnamefont {M.~A.}\ \bibnamefont
  {Troxel}} \emph {et~al.} (\bibinfo {collaboration} {Dark Energy Survey
  Collaboration}),\ }\href {https://doi.org/10.1103/PhysRevD.98.043528}
  {\bibfield  {journal} {\bibinfo  {journal} {Phys. Rev. D}\ }\textbf {\bibinfo
  {volume} {98}},\ \bibinfo {pages} {043528} (\bibinfo {year}
  {2018})}\BibitemShut {NoStop}%
\bibitem [{\citenamefont {Hikage}\ \emph {et~al.}(2019)\citenamefont {Hikage}
  \emph {et~al.}}]{Hikage2019}%
  \BibitemOpen
  \bibfield  {author} {\bibinfo {author} {\bibfnamefont {C.}~\bibnamefont
  {Hikage}} \emph {et~al.},\ }\href {https://doi.org/10.1093/pasj/psz010}
  {\bibfield  {journal} {\bibinfo  {journal} {Publications of the Astronomical
  Society of Japan}\ }\textbf {\bibinfo {volume} {71}},\ \bibinfo {pages} {43}
  (\bibinfo {year} {2019})}\BibitemShut {NoStop}%
\bibitem [{\citenamefont {{Wright, Angus H.}}\ \emph
  {et~al.}(2020)\citenamefont {{Wright, Angus H.}} \emph
  {et~al.}}]{Wright2020}%
  \BibitemOpen
  \bibfield  {author} {\bibinfo {author} {\bibnamefont {{Wright, Angus H.}}}
  \emph {et~al.},\ }\href {https://doi.org/10.1051/0004-6361/202038389}
  {\bibfield  {journal} {\bibinfo  {journal} {A\&A}\ }\textbf {\bibinfo
  {volume} {640}},\ \bibinfo {pages} {L14} (\bibinfo {year}
  {2020})}\BibitemShut {NoStop}%
\bibitem [{\citenamefont {Poulin}\ \emph {et~al.}(2022)\citenamefont {Poulin},
  \citenamefont {Bernal}, \citenamefont {Kovetz},\ and\ \citenamefont
  {Kamionkowski}}]{poulin2022sigma}%
  \BibitemOpen
  \bibfield  {author} {\bibinfo {author} {\bibfnamefont {V.}~\bibnamefont
  {Poulin}}, \bibinfo {author} {\bibfnamefont {J.~L.}\ \bibnamefont {Bernal}},
  \bibinfo {author} {\bibfnamefont {E.}~\bibnamefont {Kovetz}},\ and\ \bibinfo
  {author} {\bibfnamefont {M.}~\bibnamefont {Kamionkowski}},\ }\href@noop {}
  {\bibfield  {journal} {\bibinfo  {journal} {arXiv preprint arXiv:2209.06217}\
  } (\bibinfo {year} {2022})}\BibitemShut {NoStop}%
\bibitem [{\citenamefont {Collaboration}(2022)}]{DESY3}%
  \BibitemOpen
  \bibfield  {author} {\bibinfo {author} {\bibfnamefont {D.}~\bibnamefont
  {Collaboration}} (\bibinfo {collaboration} {DES Collaboration}),\ }\href
  {https://doi.org/10.1103/PhysRevD.105.023514} {\bibfield  {journal} {\bibinfo
   {journal} {Phys. Rev. D}\ }\textbf {\bibinfo {volume} {105}},\ \bibinfo
  {pages} {023514} (\bibinfo {year} {2022})}\BibitemShut {NoStop}%
\bibitem [{\citenamefont {Li}\ \emph {et~al.}(2023)\citenamefont {Li} \emph
  {et~al.}}]{HSCY3}%
  \BibitemOpen
  \bibfield  {author} {\bibinfo {author} {\bibfnamefont {X.}~\bibnamefont {Li}}
  \emph {et~al.},\ }\href@noop {} {\bibinfo {title} {Hyper suprime-cam year 3
  results: Cosmology from cosmic shear two-point correlation functions}}
  (\bibinfo {year} {2023}),\ \Eprint {https://arxiv.org/abs/2304.00702}
  {arXiv:2304.00702 [astro-ph.CO]} \BibitemShut {NoStop}%
\bibitem [{\citenamefont {Nunes}\ and\ \citenamefont
  {Vagnozzi}(2021)}]{Nunes:2021ipq}%
  \BibitemOpen
  \bibfield  {author} {\bibinfo {author} {\bibfnamefont {R.~C.}\ \bibnamefont
  {Nunes}}\ and\ \bibinfo {author} {\bibfnamefont {S.}~\bibnamefont
  {Vagnozzi}},\ }\href {https://doi.org/10.1093/mnras/stab1613} {\bibfield
  {journal} {\bibinfo  {journal} {Mon. Not. Roy. Astron. Soc.}\ }\textbf
  {\bibinfo {volume} {505}},\ \bibinfo {pages} {5427} (\bibinfo {year}
  {2021})},\ \Eprint {https://arxiv.org/abs/2106.01208} {arXiv:2106.01208
  [astro-ph.CO]} \BibitemShut {NoStop}%
\bibitem [{\citenamefont {Di~Valentino}\ and\ \citenamefont
  {Bridle}(2018)}]{sym10110585}%
  \BibitemOpen
  \bibfield  {author} {\bibinfo {author} {\bibfnamefont {E.}~\bibnamefont
  {Di~Valentino}}\ and\ \bibinfo {author} {\bibfnamefont {S.}~\bibnamefont
  {Bridle}},\ }\bibfield  {journal} {\bibinfo  {journal} {Symmetry}\ }\textbf
  {\bibinfo {volume} {10}},\ \href {https://doi.org/10.3390/sym10110585}
  {10.3390/sym10110585} (\bibinfo {year} {2018})\BibitemShut {NoStop}%
\bibitem [{\citenamefont {Di~Valentino}\ \emph {et~al.}(2015)\citenamefont
  {Di~Valentino}, \citenamefont {Melchiorri},\ and\ \citenamefont
  {Silk}}]{PhysRevD.92.121302}%
  \BibitemOpen
  \bibfield  {author} {\bibinfo {author} {\bibfnamefont {E.}~\bibnamefont
  {Di~Valentino}}, \bibinfo {author} {\bibfnamefont {A.}~\bibnamefont
  {Melchiorri}},\ and\ \bibinfo {author} {\bibfnamefont {J.}~\bibnamefont
  {Silk}},\ }\href {https://doi.org/10.1103/PhysRevD.92.121302} {\bibfield
  {journal} {\bibinfo  {journal} {Phys. Rev. D}\ }\textbf {\bibinfo {volume}
  {92}},\ \bibinfo {pages} {121302(R)} (\bibinfo {year} {2015})}\BibitemShut
  {NoStop}%
\bibitem [{\citenamefont {Di~Valentino}\ \emph
  {et~al.}(2020{\natexlab{a}})\citenamefont {Di~Valentino}, \citenamefont
  {Melchiorri}, \citenamefont {Mena},\ and\ \citenamefont
  {Vagnozzi}}]{DiValentino:2019ffd}%
  \BibitemOpen
  \bibfield  {author} {\bibinfo {author} {\bibfnamefont {E.}~\bibnamefont
  {Di~Valentino}}, \bibinfo {author} {\bibfnamefont {A.}~\bibnamefont
  {Melchiorri}}, \bibinfo {author} {\bibfnamefont {O.}~\bibnamefont {Mena}},\
  and\ \bibinfo {author} {\bibfnamefont {S.}~\bibnamefont {Vagnozzi}},\ }\href
  {https://doi.org/10.1016/j.dark.2020.100666} {\bibfield  {journal} {\bibinfo
  {journal} {Phys. Dark Univ.}\ }\textbf {\bibinfo {volume} {30}},\ \bibinfo
  {pages} {100666} (\bibinfo {year} {2020}{\natexlab{a}})},\ \Eprint
  {https://arxiv.org/abs/1908.04281} {arXiv:1908.04281 [astro-ph.CO]}
  \BibitemShut {NoStop}%
\bibitem [{\citenamefont {Lucca}(2021)}]{LUCCA2021100899}%
  \BibitemOpen
  \bibfield  {author} {\bibinfo {author} {\bibfnamefont {M.}~\bibnamefont
  {Lucca}},\ }\href
  {https://doi.org/https://doi.org/10.1016/j.dark.2021.100899} {\bibfield
  {journal} {\bibinfo  {journal} {Physics of the Dark Universe}\ }\textbf
  {\bibinfo {volume} {34}},\ \bibinfo {pages} {100899} (\bibinfo {year}
  {2021})}\BibitemShut {NoStop}%
\bibitem [{\citenamefont {Di~Valentino}\ \emph
  {et~al.}(2020{\natexlab{b}})\citenamefont {Di~Valentino}, \citenamefont
  {Melchiorri}, \citenamefont {Mena},\ and\ \citenamefont
  {Vagnozzi}}]{Valeentino2020}%
  \BibitemOpen
  \bibfield  {author} {\bibinfo {author} {\bibfnamefont {E.}~\bibnamefont
  {Di~Valentino}}, \bibinfo {author} {\bibfnamefont {A.}~\bibnamefont
  {Melchiorri}}, \bibinfo {author} {\bibfnamefont {O.}~\bibnamefont {Mena}},\
  and\ \bibinfo {author} {\bibfnamefont {S.}~\bibnamefont {Vagnozzi}},\ }\href
  {https://doi.org/10.1103/PhysRevD.101.063502} {\bibfield  {journal} {\bibinfo
   {journal} {Phys. Rev. D}\ }\textbf {\bibinfo {volume} {101}},\ \bibinfo
  {pages} {063502} (\bibinfo {year} {2020}{\natexlab{b}})}\BibitemShut
  {NoStop}%
\bibitem [{\citenamefont {{Planck Collaboration}}\ \emph
  {et~al.}(2016)\citenamefont {{Planck Collaboration}} \emph
  {et~al.}}]{Planck2015beyond}%
  \BibitemOpen
  \bibfield  {author} {\bibinfo {author} {\bibnamefont {{Planck
  Collaboration}}} \emph {et~al.},\ }\href
  {https://doi.org/10.1051/0004-6361/201525814} {\bibfield  {journal} {\bibinfo
   {journal} {A\&A}\ }\textbf {\bibinfo {volume} {594}},\ \bibinfo {pages}
  {A14} (\bibinfo {year} {2016})}\BibitemShut {NoStop}%
\bibitem [{\citenamefont {Alishahiha}\ \emph {et~al.}(2004)\citenamefont
  {Alishahiha}, \citenamefont {Silverstein},\ and\ \citenamefont
  {Tong}}]{Alishahiha:2004eh}%
  \BibitemOpen
  \bibfield  {author} {\bibinfo {author} {\bibfnamefont {M.}~\bibnamefont
  {Alishahiha}}, \bibinfo {author} {\bibfnamefont {E.}~\bibnamefont
  {Silverstein}},\ and\ \bibinfo {author} {\bibfnamefont {D.}~\bibnamefont
  {Tong}},\ }\href {https://doi.org/10.1103/PhysRevD.70.123505} {\bibfield
  {journal} {\bibinfo  {journal} {Phys. Rev. D}\ }\textbf {\bibinfo {volume}
  {70}},\ \bibinfo {pages} {123505} (\bibinfo {year} {2004})},\ \Eprint
  {https://arxiv.org/abs/hep-th/0404084} {arXiv:hep-th/0404084} \BibitemShut
  {NoStop}%
\bibitem [{\citenamefont {Chimento}\ \emph {et~al.}(2010)\citenamefont
  {Chimento}, \citenamefont {Lazkoz},\ and\ \citenamefont
  {Sendra}}]{Chimento2010}%
  \BibitemOpen
  \bibfield  {author} {\bibinfo {author} {\bibfnamefont {L.~P.}\ \bibnamefont
  {Chimento}}, \bibinfo {author} {\bibfnamefont {R.}~\bibnamefont {Lazkoz}},\
  and\ \bibinfo {author} {\bibfnamefont {I.}~\bibnamefont {Sendra}},\ }\href
  {https://doi.org/10.1007/s10714-009-0901-z} {\bibfield  {journal} {\bibinfo
  {journal} {General Relativity and Gravitation}\ }\textbf {\bibinfo {volume}
  {42}},\ \bibinfo {pages} {1189,1209} (\bibinfo {year} {2010})}\BibitemShut
  {NoStop}%
\bibitem [{\citenamefont {Bordemann}\ and\ \citenamefont
  {Hoppe}(1993)}]{BORDEMANN1993}%
  \BibitemOpen
  \bibfield  {author} {\bibinfo {author} {\bibfnamefont {M.}~\bibnamefont
  {Bordemann}}\ and\ \bibinfo {author} {\bibfnamefont {J.}~\bibnamefont
  {Hoppe}},\ }\href
  {https://doi.org/https://doi.org/10.1016/0370-2693(93)91002-5} {\bibfield
  {journal} {\bibinfo  {journal} {Physics Letters B}\ }\textbf {\bibinfo
  {volume} {317}},\ \bibinfo {pages} {315} (\bibinfo {year}
  {1993})}\BibitemShut {NoStop}%
\bibitem [{\citenamefont {Bordemann}\ and\ \citenamefont
  {Hoppe}(1994)}]{BORDEMANN1994}%
  \BibitemOpen
  \bibfield  {author} {\bibinfo {author} {\bibfnamefont {M.}~\bibnamefont
  {Bordemann}}\ and\ \bibinfo {author} {\bibfnamefont {J.}~\bibnamefont
  {Hoppe}},\ }\href
  {https://doi.org/https://doi.org/10.1016/0370-2693(94)90025-6} {\bibfield
  {journal} {\bibinfo  {journal} {Physics Letters B}\ }\textbf {\bibinfo
  {volume} {325}},\ \bibinfo {pages} {359} (\bibinfo {year}
  {1994})}\BibitemShut {NoStop}%
\bibitem [{\citenamefont {Ogawa}(2000)}]{Ogawa2000}%
  \BibitemOpen
  \bibfield  {author} {\bibinfo {author} {\bibfnamefont {N.}~\bibnamefont
  {Ogawa}},\ }\href {https://doi.org/10.1103/PhysRevD.62.085023} {\bibfield
  {journal} {\bibinfo  {journal} {Phys. Rev. D}\ }\textbf {\bibinfo {volume}
  {62}},\ \bibinfo {pages} {085023} (\bibinfo {year} {2000})}\BibitemShut
  {NoStop}%
\bibitem [{\citenamefont {Hu}(1998)}]{Hu_1998}%
  \BibitemOpen
  \bibfield  {author} {\bibinfo {author} {\bibfnamefont {W.}~\bibnamefont
  {Hu}},\ }\href {https://doi.org/10.1086/306274} {\bibfield  {journal}
  {\bibinfo  {journal} {The Astrophysical Journal}\ }\textbf {\bibinfo {volume}
  {506}},\ \bibinfo {pages} {485} (\bibinfo {year} {1998})}\BibitemShut
  {NoStop}%
\bibitem [{\citenamefont {Hu}(2004)}]{hu2004covariant}%
  \BibitemOpen
  \bibfield  {author} {\bibinfo {author} {\bibfnamefont {W.}~\bibnamefont
  {Hu}},\ }\href@noop {} {\bibinfo {title} {Covariant linear perturbation
  formalism}} (\bibinfo {year} {2004}),\ \Eprint
  {https://arxiv.org/abs/astro-ph/0402060} {arXiv:astro-ph/0402060 [astro-ph]}
  \BibitemShut {NoStop}%
\bibitem [{\citenamefont {Gorini}\ \emph {et~al.}(2003)\citenamefont {Gorini},
  \citenamefont {Kamenshchik},\ and\ \citenamefont
  {Moschella}}]{PhysRevD.67.063509}%
  \BibitemOpen
  \bibfield  {author} {\bibinfo {author} {\bibfnamefont {V.}~\bibnamefont
  {Gorini}}, \bibinfo {author} {\bibfnamefont {A.}~\bibnamefont
  {Kamenshchik}},\ and\ \bibinfo {author} {\bibfnamefont {U.}~\bibnamefont
  {Moschella}},\ }\href {https://doi.org/10.1103/PhysRevD.67.063509} {\bibfield
   {journal} {\bibinfo  {journal} {Phys. Rev. D}\ }\textbf {\bibinfo {volume}
  {67}},\ \bibinfo {pages} {063509} (\bibinfo {year} {2003})}\BibitemShut
  {NoStop}%
\bibitem [{\citenamefont {Bilić}\ \emph {et~al.}(2002)\citenamefont {Bilić},
  \citenamefont {Tupper},\ and\ \citenamefont {Viollier}}]{BILIC200217}%
  \BibitemOpen
  \bibfield  {author} {\bibinfo {author} {\bibfnamefont {N.}~\bibnamefont
  {Bilić}}, \bibinfo {author} {\bibfnamefont {G.}~\bibnamefont {Tupper}},\
  and\ \bibinfo {author} {\bibfnamefont {R.}~\bibnamefont {Viollier}},\ }\href
  {https://doi.org/https://doi.org/10.1016/S0370-2693(02)01716-1} {\bibfield
  {journal} {\bibinfo  {journal} {Physics Letters B}\ }\textbf {\bibinfo
  {volume} {535}},\ \bibinfo {pages} {17} (\bibinfo {year} {2002})}\BibitemShut
  {NoStop}%
\bibitem [{\citenamefont {Salahedin}\ \emph {et~al.}(2020)\citenamefont
  {Salahedin}, \citenamefont {Pazhouhesh},\ and\ \citenamefont
  {Malekjani}}]{salahedin2020}%
  \BibitemOpen
  \bibfield  {author} {\bibinfo {author} {\bibfnamefont {F.}~\bibnamefont
  {Salahedin}}, \bibinfo {author} {\bibfnamefont {R.}~\bibnamefont
  {Pazhouhesh}},\ and\ \bibinfo {author} {\bibfnamefont {M.}~\bibnamefont
  {Malekjani}},\ }\href {https://doi.org/10.1140/epjp/s13360-020-00429-1}
  {\bibfield  {journal} {\bibinfo  {journal} {The European Physical Journal
  Plus}\ }\textbf {\bibinfo {volume} {135}} (\bibinfo {year}
  {2020})}\BibitemShut {NoStop}%
\bibitem [{\citenamefont {{Adil}}\ \emph {et~al.}(2023)\citenamefont {{Adil}},
  \citenamefont {{Akarsu}}, \citenamefont {{Malekjani}}, \citenamefont
  {{Colg{\'a}in}}, \citenamefont {{Pourojaghi}}, \citenamefont {{Sen}},\ and\
  \citenamefont {{Sheikh-Jabbari}}}]{Adil2023}%
  \BibitemOpen
  \bibfield  {author} {\bibinfo {author} {\bibfnamefont {S.~A.}\ \bibnamefont
  {{Adil}}}, \bibinfo {author} {\bibfnamefont {{\"O}.}~\bibnamefont
  {{Akarsu}}}, \bibinfo {author} {\bibfnamefont {M.}~\bibnamefont
  {{Malekjani}}}, \bibinfo {author} {\bibfnamefont {E.~{\'O}.}\ \bibnamefont
  {{Colg{\'a}in}}}, \bibinfo {author} {\bibfnamefont {S.}~\bibnamefont
  {{Pourojaghi}}}, \bibinfo {author} {\bibfnamefont {A.~A.}\ \bibnamefont
  {{Sen}}},\ and\ \bibinfo {author} {\bibfnamefont {M.~M.}\ \bibnamefont
  {{Sheikh-Jabbari}}},\ }\href {https://doi.org/10.48550/arXiv.2303.06928}
  {\bibfield  {journal} {\bibinfo  {journal} {arXiv e-prints}\ ,\ \bibinfo
  {eid} {arXiv:2303.06928}} (\bibinfo {year} {2023})},\ \Eprint
  {https://arxiv.org/abs/2303.06928} {arXiv:2303.06928 [astro-ph.CO]}
  \BibitemShut {NoStop}%
\bibitem [{\citenamefont {{Lin}}\ \emph {et~al.}(2023)\citenamefont {{Lin}},
  \citenamefont {{Jain}}, \citenamefont {{Raveri}}, \citenamefont {{Baxter}},
  \citenamefont {{Chang}}, \citenamefont {{Gatti}}, \citenamefont {{Lee}},\
  and\ \citenamefont {{Muir}}}]{2023arXiv230816183L}%
  \BibitemOpen
  \bibfield  {author} {\bibinfo {author} {\bibfnamefont {M.-X.}\ \bibnamefont
  {{Lin}}}, \bibinfo {author} {\bibfnamefont {B.}~\bibnamefont {{Jain}}},
  \bibinfo {author} {\bibfnamefont {M.}~\bibnamefont {{Raveri}}}, \bibinfo
  {author} {\bibfnamefont {E.~J.}\ \bibnamefont {{Baxter}}}, \bibinfo {author}
  {\bibfnamefont {C.}~\bibnamefont {{Chang}}}, \bibinfo {author} {\bibfnamefont
  {M.}~\bibnamefont {{Gatti}}}, \bibinfo {author} {\bibfnamefont
  {S.}~\bibnamefont {{Lee}}},\ and\ \bibinfo {author} {\bibfnamefont
  {J.}~\bibnamefont {{Muir}}},\ }\href
  {https://doi.org/10.48550/arXiv.2308.16183} {\bibfield  {journal} {\bibinfo
  {journal} {arXiv e-prints}\ ,\ \bibinfo {eid} {arXiv:2308.16183}} (\bibinfo
  {year} {2023})},\ \Eprint {https://arxiv.org/abs/2308.16183}
  {arXiv:2308.16183 [astro-ph.CO]} \BibitemShut {NoStop}%
\bibitem [{\citenamefont {Smith}\ \emph {et~al.}(2003)\citenamefont {Smith},
  \citenamefont {Peacock}, \citenamefont {Jenkins}, \citenamefont {White},
  \citenamefont {Frenk}, \citenamefont {Pearce}, \citenamefont {Thomas},
  \citenamefont {Efstathiou},\ and\ \citenamefont {Couchman}}]{smith2003}%
  \BibitemOpen
  \bibfield  {author} {\bibinfo {author} {\bibfnamefont {R.~E.}\ \bibnamefont
  {Smith}}, \bibinfo {author} {\bibfnamefont {J.~A.}\ \bibnamefont {Peacock}},
  \bibinfo {author} {\bibfnamefont {A.}~\bibnamefont {Jenkins}}, \bibinfo
  {author} {\bibfnamefont {S.~D.~M.}\ \bibnamefont {White}}, \bibinfo {author}
  {\bibfnamefont {C.~S.}\ \bibnamefont {Frenk}}, \bibinfo {author}
  {\bibfnamefont {F.~R.}\ \bibnamefont {Pearce}}, \bibinfo {author}
  {\bibfnamefont {P.~A.}\ \bibnamefont {Thomas}}, \bibinfo {author}
  {\bibfnamefont {G.}~\bibnamefont {Efstathiou}},\ and\ \bibinfo {author}
  {\bibfnamefont {H.~M.~P.}\ \bibnamefont {Couchman}},\ }\href
  {https://doi.org/10.1046/j.1365-8711.2003.06503.x} {\bibfield  {journal}
  {\bibinfo  {journal} {Monthly Notices of the Royal Astronomical Society}\
  }\textbf {\bibinfo {volume} {341}},\ \bibinfo {pages} {1311} (\bibinfo {year}
  {2003})}\BibitemShut {NoStop}%
\bibitem [{\citenamefont {Mead}\ \emph {et~al.}(2016)\citenamefont {Mead},
  \citenamefont {Heymans}, \citenamefont {Lombriser}, \citenamefont {Peacock},
  \citenamefont {Steele},\ and\ \citenamefont {Winther}}]{Mead2016}%
  \BibitemOpen
  \bibfield  {author} {\bibinfo {author} {\bibfnamefont {A.~J.}\ \bibnamefont
  {Mead}}, \bibinfo {author} {\bibfnamefont {C.}~\bibnamefont {Heymans}},
  \bibinfo {author} {\bibfnamefont {L.}~\bibnamefont {Lombriser}}, \bibinfo
  {author} {\bibfnamefont {J.~A.}\ \bibnamefont {Peacock}}, \bibinfo {author}
  {\bibfnamefont {O.~I.}\ \bibnamefont {Steele}},\ and\ \bibinfo {author}
  {\bibfnamefont {H.~A.}\ \bibnamefont {Winther}},\ }\href
  {https://doi.org/10.1093/mnras/stw681} {\bibfield  {journal} {\bibinfo
  {journal} {Monthly Notices of the Royal Astronomical Society}\ }\textbf
  {\bibinfo {volume} {459}},\ \bibinfo {pages} {1468} (\bibinfo {year}
  {2016})}\BibitemShut {NoStop}%
\bibitem [{\citenamefont {{Kuijken, K.}}\ \emph {et~al.}(2019)\citenamefont
  {{Kuijken, K.}} \emph {et~al.}}]{Kuijken2019}%
  \BibitemOpen
  \bibfield  {author} {\bibinfo {author} {\bibnamefont {{Kuijken, K.}}} \emph
  {et~al.},\ }\href {https://doi.org/10.1051/0004-6361/201834918} {\bibfield
  {journal} {\bibinfo  {journal} {A\&A}\ }\textbf {\bibinfo {volume} {625}},\
  \bibinfo {pages} {A2} (\bibinfo {year} {2019})}\BibitemShut {NoStop}%
\bibitem [{\citenamefont {Alam}\ \emph {et~al.}(2017)\citenamefont {Alam} \emph
  {et~al.}}]{Alam2017}%
  \BibitemOpen
  \bibfield  {author} {\bibinfo {author} {\bibfnamefont {S.}~\bibnamefont
  {Alam}} \emph {et~al.},\ }\href {https://doi.org/10.1093/mnras/stx721}
  {\bibfield  {journal} {\bibinfo  {journal} {Monthly Notices of the Royal
  Astronomical Society}\ }\textbf {\bibinfo {volume} {470}},\ \bibinfo {pages}
  {2617} (\bibinfo {year} {2017})}\BibitemShut {NoStop}%
\bibitem [{\citenamefont {Blas}\ \emph {et~al.}(2011)\citenamefont {Blas},
  \citenamefont {Lesgourgues},\ and\ \citenamefont {Tram}}]{Diego_Blas_2011}%
  \BibitemOpen
  \bibfield  {author} {\bibinfo {author} {\bibfnamefont {D.}~\bibnamefont
  {Blas}}, \bibinfo {author} {\bibfnamefont {J.}~\bibnamefont {Lesgourgues}},\
  and\ \bibinfo {author} {\bibfnamefont {T.}~\bibnamefont {Tram}},\ }\href
  {https://doi.org/10.1088/1475-7516/2011/07/034} {\bibfield  {journal}
  {\bibinfo  {journal} {Journal of Cosmology and Astroparticle Physics}\
  }\textbf {\bibinfo {volume} {2011}}\bibinfo  {number} { (07)},\ \bibinfo
  {pages} {034}}\BibitemShut {NoStop}%
\bibitem [{Note1()}]{Note1}%
  \BibitemOpen
\bibfield  {number} {  }\bibinfo {note} {\protect \url
  {https://lesgourg.github.io/class_public/class.html}}\BibitemShut {NoStop}%
\bibitem [{\citenamefont {Brinckmann}\ and\ \citenamefont
  {Lesgourgues}(2019)}]{Thejs2019}%
  \BibitemOpen
  \bibfield  {author} {\bibinfo {author} {\bibfnamefont {T.}~\bibnamefont
  {Brinckmann}}\ and\ \bibinfo {author} {\bibfnamefont {J.}~\bibnamefont
  {Lesgourgues}},\ }\href
  {https://doi.org/https://doi.org/10.1016/j.dark.2018.100260} {\bibfield
  {journal} {\bibinfo  {journal} {Physics of the Dark Universe}\ }\textbf
  {\bibinfo {volume} {24}},\ \bibinfo {pages} {100260} (\bibinfo {year}
  {2019})}\BibitemShut {NoStop}%
\bibitem [{\citenamefont {Audren}\ \emph {et~al.}(2013)\citenamefont {Audren},
  \citenamefont {Lesgourgues}, \citenamefont {Benabed},\ and\ \citenamefont
  {Prunet}}]{Audren_2013}%
  \BibitemOpen
  \bibfield  {author} {\bibinfo {author} {\bibfnamefont {B.}~\bibnamefont
  {Audren}}, \bibinfo {author} {\bibfnamefont {J.}~\bibnamefont {Lesgourgues}},
  \bibinfo {author} {\bibfnamefont {K.}~\bibnamefont {Benabed}},\ and\ \bibinfo
  {author} {\bibfnamefont {S.}~\bibnamefont {Prunet}},\ }\href
  {https://doi.org/10.1088/1475-7516/2013/02/001} {\bibfield  {journal}
  {\bibinfo  {journal} {Journal of Cosmology and Astroparticle Physics}\
  }\textbf {\bibinfo {volume} {2013}}\bibinfo  {number} { (02)},\ \bibinfo
  {pages} {001}}\BibitemShut {NoStop}%
\bibitem [{Note2()}]{Note2}%
  \BibitemOpen
\bibfield  {number} {  }\bibinfo {note} {\protect \url
  {https://baudren.github.io/montepython.html}}\BibitemShut {NoStop}%
\bibitem [{\citenamefont {{Planck
  Collaboration}}(2020{\natexlab{b}})}]{Planck2018likelihood}%
  \BibitemOpen
  \bibfield  {author} {\bibinfo {author} {\bibnamefont {{Planck
  Collaboration}}},\ }\href {https://doi.org/10.1051/0004-6361/201936386}
  {\bibfield  {journal} {\bibinfo  {journal} {A\&A}\ }\textbf {\bibinfo
  {volume} {641}},\ \bibinfo {pages} {A5} (\bibinfo {year}
  {2020}{\natexlab{b}})}\BibitemShut {NoStop}%
\bibitem [{\citenamefont {{Joachimi, B.}}\ \emph {et~al.}(2021)\citenamefont
  {{Joachimi, B.}} \emph {et~al.}}]{Joachimi2021}%
  \BibitemOpen
  \bibfield  {author} {\bibinfo {author} {\bibnamefont {{Joachimi, B.}}} \emph
  {et~al.},\ }\href {https://doi.org/10.1051/0004-6361/202038831} {\bibfield
  {journal} {\bibinfo  {journal} {A\&A}\ }\textbf {\bibinfo {volume} {646}},\
  \bibinfo {pages} {A129} (\bibinfo {year} {2021})}\BibitemShut {NoStop}%
\bibitem [{\citenamefont {Zuntz}\ \emph {et~al.}(2015)\citenamefont {Zuntz},
  \citenamefont {Paterno}, \citenamefont {Jennings}, \citenamefont {Rudd},
  \citenamefont {Manzotti}, \citenamefont {Dodelson}, \citenamefont {Bridle},
  \citenamefont {Sehrish},\ and\ \citenamefont {Kowalkowski}}]{ZUNTZ201545}%
  \BibitemOpen
  \bibfield  {author} {\bibinfo {author} {\bibfnamefont {J.}~\bibnamefont
  {Zuntz}}, \bibinfo {author} {\bibfnamefont {M.}~\bibnamefont {Paterno}},
  \bibinfo {author} {\bibfnamefont {E.}~\bibnamefont {Jennings}}, \bibinfo
  {author} {\bibfnamefont {D.}~\bibnamefont {Rudd}}, \bibinfo {author}
  {\bibfnamefont {A.}~\bibnamefont {Manzotti}}, \bibinfo {author}
  {\bibfnamefont {S.}~\bibnamefont {Dodelson}}, \bibinfo {author}
  {\bibfnamefont {S.}~\bibnamefont {Bridle}}, \bibinfo {author} {\bibfnamefont
  {S.}~\bibnamefont {Sehrish}},\ and\ \bibinfo {author} {\bibfnamefont
  {J.}~\bibnamefont {Kowalkowski}},\ }\href
  {https://doi.org/https://doi.org/10.1016/j.ascom.2015.05.005} {\bibfield
  {journal} {\bibinfo  {journal} {Astronomy and Computing}\ }\textbf {\bibinfo
  {volume} {12}},\ \bibinfo {pages} {45} (\bibinfo {year} {2015})}\BibitemShut
  {NoStop}%
\bibitem [{\citenamefont {{van Dokkum}}\ \emph {et~al.}(2018)\citenamefont
  {{van Dokkum}} \emph {et~al.}}]{Dokkum2018}%
  \BibitemOpen
  \bibfield  {author} {\bibinfo {author} {\bibfnamefont {P.}~\bibnamefont {{van
  Dokkum}}} \emph {et~al.},\ }\href {https://doi.org/10.1038/nature25767}
  {\bibfield  {journal} {\bibinfo  {journal} {\nat}\ }\textbf {\bibinfo
  {volume} {555}},\ \bibinfo {pages} {629} (\bibinfo {year} {2018})},\ \Eprint
  {https://arxiv.org/abs/1803.10237} {arXiv:1803.10237 [astro-ph.GA]}
  \BibitemShut {NoStop}%
\bibitem [{\citenamefont {{Ogiya}}\ \emph {et~al.}(2022)\citenamefont
  {{Ogiya}}, \citenamefont {{van den Bosch}}, \citenamefont {{Burkert}},\ and\
  \citenamefont {{Kang}}}]{Ogiya2022}%
  \BibitemOpen
  \bibfield  {author} {\bibinfo {author} {\bibfnamefont {G.}~\bibnamefont
  {{Ogiya}}}, \bibinfo {author} {\bibfnamefont {F.~C.}\ \bibnamefont {{van den
  Bosch}}}, \bibinfo {author} {\bibfnamefont {A.}~\bibnamefont {{Burkert}}},\
  and\ \bibinfo {author} {\bibfnamefont {X.}~\bibnamefont {{Kang}}},\ }\href
  {https://doi.org/10.3847/2041-8213/aca2a7} {\bibfield  {journal} {\bibinfo
  {journal} {The Astrophysical Journal Letters}\ }\textbf {\bibinfo {volume}
  {940}},\ \bibinfo {pages} {L46} (\bibinfo {year} {2022})}\BibitemShut
  {NoStop}%
\bibitem [{\citenamefont {Kase}\ \emph {et~al.}(2007)\citenamefont {Kase},
  \citenamefont {Makino},\ and\ \citenamefont {Funato}}]{Kase2007}%
  \BibitemOpen
  \bibfield  {author} {\bibinfo {author} {\bibfnamefont {H.}~\bibnamefont
  {Kase}}, \bibinfo {author} {\bibfnamefont {J.}~\bibnamefont {Makino}},\ and\
  \bibinfo {author} {\bibfnamefont {Y.}~\bibnamefont {Funato}},\ }\href
  {https://doi.org/10.1093/pasj/59.6.1071} {\bibfield  {journal} {\bibinfo
  {journal} {Publications of the Astronomical Society of Japan}\ }\textbf
  {\bibinfo {volume} {59}},\ \bibinfo {pages} {1071} (\bibinfo {year}
  {2007})}\BibitemShut {NoStop}%
\end{thebibliography}%
	
\end{document}